\documentclass[%
reprint,
superscriptaddress,
amsmath,amssymb,
aps,
]{revtex4-2}

\usepackage{graphicx}
\usepackage{dcolumn}
\usepackage{bm}
\usepackage{cleveref,braket,xcolor,subcaption}
\usepackage{amsmath}
\DeclareMathOperator{\tr}{tr}

\begin{document}


\title{Neural Networks for Quantum Inverse Problems}

\author{Ningping Cao}
\affiliation{Department of Mathematics $\&$ Statistics, University of Guelph, Guelph N1G 2W1, Ontario, Canada}
\affiliation{Institute for Quantum Computing, University of Waterloo, Waterloo N2L 3G1, Ontario, Canada}

\author{Jie Xie}
\affiliation{National Laboratory of Solid State Microstructures, College of Engineering and Applied Sciences and School of Physics, Nanjing University, Nanjing 210093, China}
\affiliation{Collaborative Innovation Center of Advanced Microstructures, Nanjing University, Nanjing 210093, China}

\author{Aonan Zhang}
\affiliation{National Laboratory of Solid State Microstructures, College of Engineering and Applied Sciences and School of Physics, Nanjing University, Nanjing 210093, China}
\affiliation{Collaborative Innovation Center of Advanced Microstructures, Nanjing University, Nanjing 210093, China}

\author{Shi-Yao Hou}%
\affiliation{College of Physics and Electronic Engineering, Center for Computational Sciences,  Sichuan Normal University, Chengdu 610068, People's Republic of China}%

\author{Lijian Zhang}
\email[]{lijian.zhang@nju.edu.cn}
\affiliation{National Laboratory of Solid State Microstructures, College of Engineering and Applied Sciences and School of Physics, Nanjing University, Nanjing 210093, China}
\affiliation{Collaborative Innovation Center of Advanced Microstructures, Nanjing University, Nanjing 210093, China}

\author{Bei Zeng}
\email[]{zengb@ust.hk}
\affiliation{Department of Physics, The Hong Kong University of Science and Technology, Clear Water Bay, Kowloon, Hong Kong, China}


\date{\today}

\begin{abstract}
Quantum Inverse Problem (QIP) is the problem of estimating an unknown quantum system $\rho$ from a set of measurements, whereas the classical counterpart is the Inverse Problem of estimating a distribution from a set of observations.
In this paper, we present a neural network based method for QIPs, which has been widely explored for its classical counterpart. The proposed method utilizes the quantum-ness of the QIPs and takes advantage of the computational power of neural networks to achieve higher efficiency for the quantum state estimation.
We test the method on the problem of Maximum Entropy Estimation of an unknown state $\rho$ from partial information. Our method yields high fidelity, efficiency and robustness for both numerical experiments and quantum optical experiments.
\end{abstract}

\maketitle

\section{Introduction}

Learning quantum states is an essential task in quantum information processing~\cite{rocchetto2019experimental,torlai2018neural,cramer2010efficient}. 
Normally, performing measurements on a quantum system, getting readouts, and reconstructing the quantum states is the way to study the corresponding systems.
In general this process can be written as
$\vec c = \mathcal{A}(\rho)+\vec\epsilon$,
where $\rho$ is the quantum state of the system, $\mathcal{A}$ is a function of $\rho$,
and $\vec{c}$ is expectation values obtained from the measurements that are specified by the function $\mathcal{A}$.
$\vec\epsilon$ is the noise vector that subjects to a noise distribution $\pi_\text{noise}$.
If one knows the state $\rho$, then $\vec c$ can be obtained by getting the measurements specified by $\mathcal{A}$, and we call this process the quantum forward problem; the opposite direction is then called the Quantum Inverse Problem (QIP).


Despite of the contexts of quantum systems, when the operator $\rho$ is a diagonal matrix, i.e. a classical probability distribution, the QIP reduces to its classical counterpart, i.e. the classical Inverse Problem (IP), which is known as one of the most important mathematical problems since it tells parameters that are not directly measurable, a situation widely applicable in science and technology.
From the analytic prospective, it is the problem of recovering system parameters (elements of diagonal $\rho$) from observables $\vec c$ according to a certain model; from Bayesian inference viewpoint, the goal is to recover the classical probability distribution of a diagonal $\rho$ subject to the given $\vec c$~\cite{adler2017solving}.

IP has been studied for a long time in statistical inference and statistical learning~\cite{tarantola2005inverse,bal2012introduction}.
One of the difficulties is that IPs are often \textit{ill-posted}. That means the solution may not exist, not unique when exists, or instable in the sense that a small change in the input may cause a large change to the output.
Deep neural networks are state-of-art tools for solving IP~\cite{adler2017solving,li2020nett,arridge2019solving,genzel2020solving,lucas2018using,adler2018deep,mukherjee2020learned}, sevral of them are focused on the ill-posed cases~\cite{adler2017solving,li2020nett}.
These proposals have been broadly applied to biomedical imaging~\cite{ronneberger2015u,jin2017deep,arridge2019solving,senouf2019self,prato2008inverse}, geophysics~\cite{tarantola1982inverse}, optical physics~\cite{pilozzi2018machine}, and engineering sciences~\cite{cherubini2005inversion} etc.

\begin{figure}[t]
\centering
\includegraphics[width = .9\linewidth]{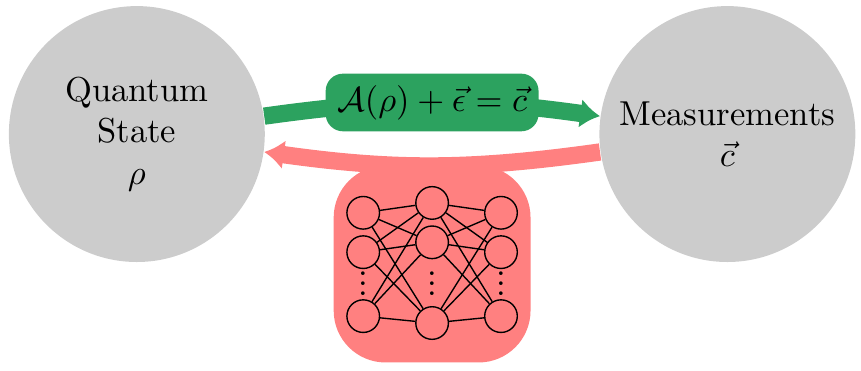}
\caption{\textbf{The schematic diagram of the general method}: the green arrow depicts the quantum forward model; the red arrow is the quantum inverse problem. We tackle the problem by supervised learning a deep neural network.}\label{fig:method}
\end{figure}

When $\mathcal{A}$ is the function of directly mapping $\rho$ to a complete measurements $\vec c$, the QIP corresponds to the so-called Quantum State Tomography.
The standard tomography process requires $d^{2} - 1$ measurements from linearly independent observables, where $d$ is the dimension of Hilbert space~\cite{nielsen2002quantum}. It is highly resource demanding since the number of measurements grows exponentially with the number of qubits, and made it not practical even in NISQ era.
The goal of QIPs in general, from statistical inference perspective, are to discover the quantum distributions based on observed $\vec c$, then give an estimation based on the distribution.
Quantum probability (non-diagonal $\rho$) is a noncommutative probability on von Neumann algebra, while classical probability is the commutative special case (its von Neumann algebra is abelian)~\cite{redei2007quantum}.
This makes quantum probability distributions even harder to learn.
Deep neural networks have great potential to give sound estimations in such difficult scenarios.
In this paper, we develop a method for QIPs using deep neural networks, it depicts in~\Cref{fig:method}.

After properly parametrize the quantum state $\rho$, deep neural networks offer a great opportunity to fulfill this task by implementing supervised learning techniques.
The common concerns of applying supervised learning techniques are training data availability and the training data distribution.
In our case, the knowledge from quantum physics pitches in, provides ways of parametrize states, and physical models for the Quantum Forward Problem (QFP).
It can generate abundant training and testing data; It also reveals the landscape of the QIP, which can determining the training data sampling and distribution.

We consider incomplete measurements to demonstrate our deep learning method since the complete measurements $d^2-1$ are not practical. 
A incomplete set of measurements $\vec c$, refer to as partial information, usually does not reveal all information of the system. A given $\vec c$ is normally corresponding to multiple preimages $\rho$.
However, with appropriate prior information, the partial information could have a almost (with probability one) bijective relation with the quantum system.
For example, based on the knowledge that the unknown quantum state has low rank, compressed sensing~\cite{gross2010quantum,flammia2012quantum,riofrio2017experimental} can reconstruct the state with Pauli operators way less than $d^2-1$;
A small number of measurements can predict many properties in a quantum system~\cite{huang2020predicting};
With certain assumptions, one eigenstate can encode all the information in a Hamiltonian~\cite{garrison2018does,qi2019determining,hou2020determining}.
That means the measurements of one eigenstates could help to recover the system Hamiltonian as well as the measured eigenstate.
Though the mathematical connection $\mathcal{A}$ between $\rho$ and expectation values $\vec c$ is clear, it is nontrivial to come up with effective schemes to solve the QIP while efficiently utilize the prior information~\cite{flammia2012quantum}.

Our method are particular good for these problems with prior information.
The almost bijective relation guaranteed by prior information of the system restricts the corresponding QIP to be likely well-posted.
The method could easily and effectively utilize the prior information of a regrading quantum system, because the information can easily embed in the QFP.
The design of specific steps for a giving QIP is intuitive, as well as the network training. Our framework not only can be applied to solve the problem, it could also offer initial values for other
Monte Carlo based methods. 
We test our scheme on the task of giving Maximum Entropy Estimation (MEE) from partial information. The method demonstrates high fidelity and extraordinary efficiency on both numerical data and optical experimental data. It also menifests ability to tolerant experimental noise.

This paper is organized in the following way: Section II presents and discusses the deep learning method; Section III demonstrates the example that giving the MEE $\rho_\text{MEE}$ of an unknown state $\rho$ from partial information in details, supported by both numerical and quantum optical experiment results; Section IV then follows with discussion and outlook. 


\section{The neural networks method}

For a quantum physical system, the underlying mechanism is governing by the physical model $\mathcal{A}_{\mathbf{F}}: X \to Y$ such that
\begin{equation}\label{eq:model}
\vec c = \mathcal{A}_{\mathbf{F}}(\rho) + \vec \epsilon,
\end{equation}
where $\rho \in X$ is the density matrix, $\vec \epsilon \in Y$ is the noise vector (subject to a noise distribution $\pi_\text{noise}$), and $\vec c \in Y$ is the expectation values of measuring a fixed set of observables $\mathbf{F} = \{F_1,\cdots,F_m\}$. $X$ is the interested set of $d\times d$ density matrices according to given prior information. For example, if we know the states $\rho$ are pure states, $X$ is then the set of rank-$1$ density matrices; $X$ is the set of all density matrices if no prior information is provided.
$Y$ is the vector space $\mathbb{R}^m$.
The physical model $\mathcal{A}_{\mathbf{F}}$ is always associates with a set $\mathbf{F}$ of observables to determine which space it maps into. For simplicity, we will use the notation $\mathcal{A}$ instead.
Giving $\rho$ to get $\vec c$ is the Quantum Forward Problem (QFP), the reverse direction we call it the Quantum Inverse Problem (QIP).

When the measurements are not complete -- that is the number $m$ of observables are not enough for uniquely determine the system, the quantum inverse problem does not process a unique solution.
Prior information of the interested system decreases the degree of freedom.
Under these prior information, a model $\mathcal{A}$ is actually a bijective or almost bijective.
That means the system can be reconstructed with fewer measurements if the prior information has been appropriately used.
However, the challenge of how to efficiently encode the prior information into a practical scheme is demanding~\cite{flammia2012quantum}.
This turns out to be a blessing for our framework, it is straightforward to utilize such information.
This will be explained later.

The key observation is that the forward problem is almost always simpler than the inverse. For example, reconstructing a object from its projections is substantially more challenging than making the projections of a known object.
Especially, in quantum information theory, the forward problem is usually clear thanks to the study of quantum physics.
Knowing the information of a quantum system, such as its Hamiltonian or state (density matrix), the measurement outcomes of this system according to a fix set $\mathbf{F}$ are predictable. Comparing to the task of reconstructing information of the system from its measurements, the forward direction is significantly easier.

We take advantage of the complexity difference between the two directions, use the easier direction to help deal with the difficult side.
Supervised learning is the relatively mature branch of machine learning techniques that finding the model between input and output pairs data with labeled examples.
On the contrary, unsupervised learning techniques are normally used while lack of access to training data or having trouble with labeling data~\cite{srivastava2015unsupervised}.
In our problem, data resource for supervised learning is guaranteed.
The QFP contributes to generate training and testing data for supervised learning.
The next problem is training data distribution.
The forward model $\mathcal{A}$ contains the information about the landscape of $\vec c$ according to $\rho$. This information guides the training data sampling process, largely determine the training data distribution.

From another perspective, QIPs are also regression problems to fit given data pairs.
Neural Networks are extremely versatile tools for regression problem.
Even simple NNs only with one hidden layer are very expensive. With nonlinear activation functions between neurons, these ``vanilla'' NNs can represent arbitrary functions~\cite{cybenko1989approximation}.
Traditionally, regression problems require deliberately chosen techniques to achieve better performance.
However, Neural Networks are extremely flexible, they automatic adapt themselves to different type of regression techniques according to the particular scenario. This made NN a convenient tool for solving various QIPs. 

Before implementation, we need to parametrize density matrices $\rho$.
The parametrization function $P: X' \to X$ is a bijection,
\begin{equation}\label{eq:para}
P(\vec a) = \rho,
\end{equation}
where $X'$ is a vector space. The choice of $P$ is based on $X$. 
For example, if the set $X$ is the Gibbs states of a class of Hamiltonian, $P$ could be the map from Hamiltonian parameters to the Gibbs states.

The training data set denotes as
\[\mathbf{T} = \{(\vec c_\text{train}, \vec a_\text{train})| \vec c_\text{train} =  \mathcal{A}(\vec a_\text{train}) + \vec \epsilon, \vec a_\text{train} \in X'_\text{train} \},
\]
where $X'_\text{train} \subset X'$ is the finite set of sampled parameters $\vec a_\text{train}$. The training data naturally implants the prior information contained in $\mathcal{A}$.

After data preparation, the network can be trained.
A neural network is a tunable function $\Phi_\text{NN}: Y \to X'$. The training process is using train algorithms to tune the parameters embedded in $\Phi_\text{NN}$ to minimize the distances between the NN output and desired output according to the chosen loss function, i.e. minimizing
\begin{equation}\label{eq:loss}
L(\vec a_\text{train}, \Phi_\text{NN}(\vec c_\text{train})),
\end{equation}
where $L: X'\times X' \to \mathbb{R}$ is the loss function.
It is chosen to reverberate the parametrization $P$.
The goal is to bring $\rho_1 $ and $\rho_2$ closer by minimizing $L(\vec a_1,\vec a_2)$.
$L$ could be Mean Square Error or Mean Absolute Error if $\vec a$s need to be precisely approached on magnitudes.
If $P$ is more focus on the direction of $\vec a$s, a loss function minimizes angle (e.g. Cosine Similarity) will be a better choice.
$L$ can also be a type of entropy when $\vec a$s are probability distributions. The choice of $P$ and $L$, the training date set $\mathbf{T}$ all reflect prior information of the problem.

For testing data generated by the QFP, comparing the ideal $\rho$ and the estimated $\rho_\text{est}$ can tell us the accuracy of the estimation.
A reasonable question to ask is, given a data $\vec c_\text{unk}$ with an unknown preimage, how can we know the NN estimation $\rho_\text{est}$ is acceptable.
It turns out that QFP can serve as the mechanism of examining the estimation that come out from a trained NN.
Choosing a metric $f$ in $Y$, one can compare $\vec c_\text{unk}$ and the image of NN output,
\begin{equation}\label{eq:exam}
f(\vec c_\text{unk}, \mathcal{A}\circ P\circ \Phi_\text{NN}(\vec c_\text{unk})).
\end{equation}
Ideally, we want $\vec c_\text{unk}$ and $(\mathcal{A}\circ P\circ \Phi_\text{NN})(\vec c_\text{unk})$ to be identical, but numerical errors are inevitable in reality.
Bounding the value of~\cref{eq:exam} can bound the confidence of $\rho_\text{est}$.





In the next section, we will provide an example to demonstrate our method. The task is to give an Maximum Entropy Estimation (MEE) based on noiseless partial information of an unknown state.
The network takes incomplete measurements of the unknown quantum state $\rho$ and returns the MEE of $\rho$.
Comparing to other algorithm, our method shows extraordinary efficiency without sacrifice much accuracy. It also shows a great ability of experimental error tolerance.

\section{Learning Maximum Entropy Estimation from partial information}

Maximum entropy inference is believed to be the best estimation to present the current knowledge 
when only part of information about the system is provided~\cite{jaynes1957information,wichmann1963density}.
The entropy is mostly Shanon Entropy in classical physics and engineering, and is von Neumann Entropy for the quantum counterpart.

In quantum system, given the set of incomplete measurement expectation values 
$\{c_i | c_i = tr(\rho F_i), F_i \in \mathbf{F}\}$
of an unknown state $\rho$ for a fixed set of observables $\mathbf{F}$, there may exist more then one quantum state has the same measurement outcomes.
The incompleteness means that the measurements are not enough for a full tomography ($m< d^2-1$).
Denote the set of state as

\[\mathbf{P} = \{\rho^*| \tr(\rho^*F_i) = \tr(\rho F_i),  \forall F_i \in \mathbf{F}\}. \]

The unknown state $\rho$ is one of the elements in $\mathbf{P}$.
The Maximum Entropy Estimation $\rho_\text{MEE}$ of $\rho$ can be represented as a thermal state
\begin{equation}
\rho_\text{MEE} = \frac{\exp(\beta \sum_i a_i F_i)}{\tr[\exp(\beta \sum_i a_i F_i)]},\label{eq:mee}
\end{equation}
where $\beta$ is the reciprocal temperature of the system and $a_i$'s are real coefficients~\cite{niekamp2013computing,chen2015discontinuity,rodman2016continuity}. At the mean time, $\rho_\text{MEE}$ should satisfy that it has the same measurement outcomes when measuring the same set of operator $\mathbf{F}$ (i.e. $\rho_\text{MEE}\in \mathbf{P}$). The thermal representation is unique~\cite{niekamp2013computing}.
The measurement results 
$\vec c = (c_1,\cdots,c_m)$ where $c_i = \tr(\rho F_i), F_i \in \mathbf{F}$, therefore, possess a one-to-one correspondence with its MEE $\rho_\text{MEE}$.

An interesting special case is that $\mathbf{P}$ only has one element, then $\rho = \rho_\text{MEE}$. A well-studied example of such case is when $\rho$ is an unique ground state (UGS) of $H = -\sum_i a_i F_i$, where $F_i \in \mathbf{F}$ and $a_i\in \mathbb{R}$~\cite{xin2017quantum,karuvade2019uniquely,PhysRevLett.125.150401}.
That means, if $\rho$ is an UGS of $H$, $\rho_\text{MEE}$ not only has an one-to-one correspondence with $\vec c$, it is also the actual state $\rho$.



\begin{figure*}[t]
	\subcaptionbox{Training and testing data generation\label{fig:data_gen}}
	{\includegraphics[width = 0.45\linewidth]{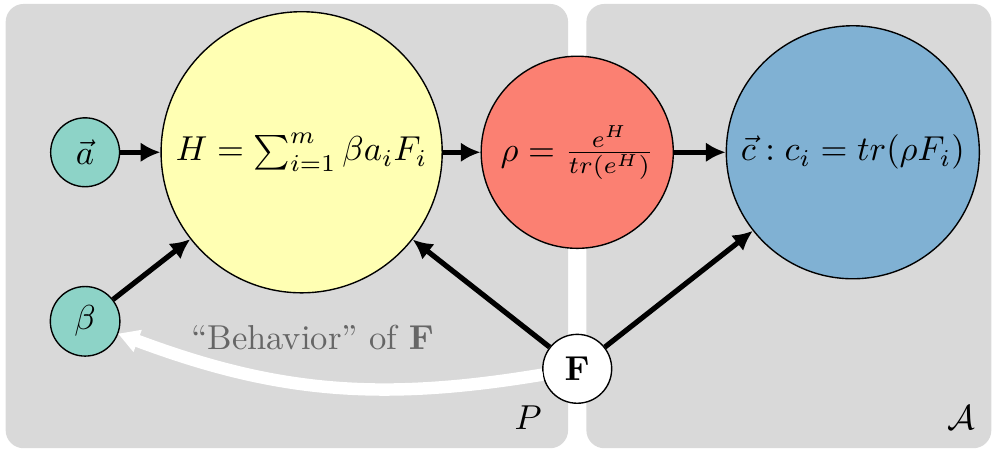}}
	\subcaptionbox{The process\label{fig:network}}
	{\includegraphics[width = 0.53\linewidth]{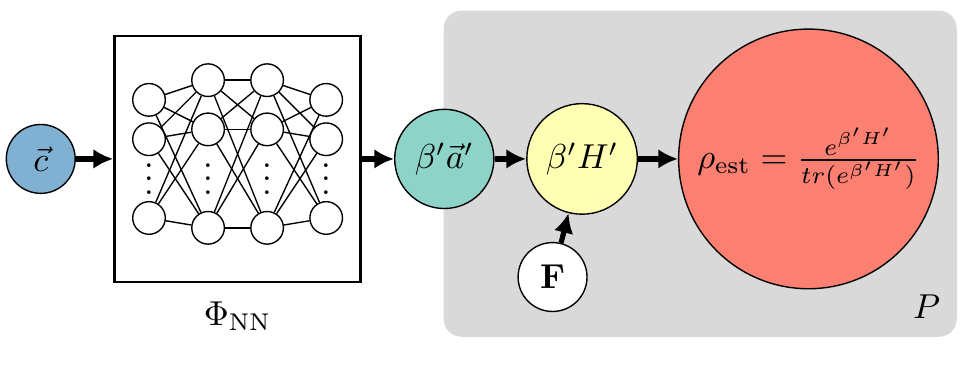}}
\caption{(a) $\beta$ and $\vec a$ are randomly generated. The hermitian operator set $\mathbf{F}$ associate with the forward model $\mathcal{A}$ provides information for the distribution of $\beta$. Generated pairs of $(\vec c, \beta\vec a)$ are training data: $\vec c$ is the input of the neural network, $\beta\vec a$ is the idea output. The left block is the parametrization function $P$, the right block is the foreword model $\mathcal{A}$ for this problem.
(b) the trained network $\Phi_\text{NN}$ is the approximation of $P^{-1}\circ\mathcal{A}^{-1}$, can produce a parameter estimation $\beta'\vec a'$ according to an input $\vec c$. The right block is the parameterization function $P$ that maps estimated parameters $\beta'\vec a'$ to the estimation of MEE $\rho_\text{est}$
}
\end{figure*}

In this particular QIP,
the parametrization function $P$ and the forward model $\mathcal{A}$ are as follows:
\begin{equation}
P: \beta\vec a \to \rho_\text{MEE}
\label{eq:para_th}
\end{equation}
\[
\mathcal{A}: \rho_\text{MEE} \to \vec c
\]
where $\vec a = (a_1,\cdots,a_m)$ and $\rho_\text{MEE}$ is defined in~\Cref{eq:mee}.
Supervised learning can train a network $\Phi_\text{NN}$ to approach the inverse function
\begin{equation}
P^{-1}\circ\mathcal{A}^{-1}: \vec c \to\beta \vec a.
\label{eq:inver}
\end{equation}
More specifically, as shown in~\Cref{fig:data_gen}, we randomly generate many $\beta$'s and $\vec a$'s, achieving corresponding measurement results $\vec c$. These pairs of $(\vec c,\beta \vec a)$ are used as training data for the neural network.
The trained network $\Phi_\text{NN}$ is the approximation of the function $P^{-1}\circ\mathcal{A}^{-1}$.
The estimation of MEE
\[
\rho_{\text{est}} = \frac{\exp(\sum_i \beta' a_i'F_i)}{\tr[\exp(\sum_i \beta' a_i'F_i)]}
\]
follows through from $P \circ\Phi_\text{NN}(\vec c)$.
To be noticed that, when the unknown state $\rho$ is not a thermal state, the operator $-H' = -\sum_i a_i' F_i$ is not necessarily the real Hamiltonian of the system. We call $H'$ a \textit{pseudo Hamiltonian}.

We test our method numerically with two systems: 1) $\mathbf{F}$ has three 64 by 64 random generated hermitian operators; 2) the 5-qubit one-dimension lattice, $\mathbf{F} = \{\sigma_a^{(i)}\otimes \sigma_b^{(i+1)}| \sigma_a, \sigma_b \in \mathcal{P}, 1\le i \le 4, a+b \neq 0\}$ where $\mathcal{P} = \{\sigma_0 = I, \sigma_1 = \sigma_x, \sigma_2 = \sigma_y, \sigma_3 = \sigma_z\}$ is the set of Pauli operators with the 2 by 2 identity. The upper index $i$ indicates the qubit of which the operator acts on. Moreover, we exam the method with experimental data of an optical set-up, which are derived from unique ground states associate with fixed hermitian operator sets. Therefore the MEE estimations $\rho_\text{est}$ are also the estimation of the true states measured in our experiments. 

\subsection{Data preparation and network training}\label{sec:data_pre}


Training data preparation is the key to supervised learning since the learning outcome depends heavily on the training data set.

The data generating procedure is shown in~\Cref{fig:data_gen}. Parameter $\vec a$ is drawn from normal $\mathcal{N}(0,1)$ distribution then normalized. The ``coldness'' $\beta$ is randomly sampled from $(0,100]$.
Generally, when $\beta$ reaches $\sim 30$, thermal states are almost pure. Here we allow $\beta$ goes to 100 for some extreme cases.
The distribution of $\beta$ in the whole training data set is critical in this process, we will discuss it in depth later. Parameter $\vec a$ together with the fixed set of operators $\mathbf{F}$ set up the pseudo Hamiltonian $H = \sum_i \beta a_i F_i$. The measurement results $\vec c = (\tr(\rho F_1),\cdots, \tr(\rho F_m))$ come from trace the product of $\rho = \exp(H)/\tr[\exp(H)]$ and operator $F_i$'s. Every pair of $\beta\vec a$ and $\vec c$ counts for a pair of training data.


It turns out that the distribution of $\beta$ in the training data set is the key to our problem. By data distribution of $\beta$, we meant the proportion of $\beta$ picked in a given interval $I$ to the amount of data in the whole training data set.
Intuitively, the network should be trained with more data in the place where the function changes more rapidly.
To be more specific, the network should see more data on where the small change of $\beta$ cause big change on $\rho$ then on $\vec c$ in the relative sense.
Despite the matrix exponential function, the property clearly also depends on $\mathbf{F}$. 
Luckily enough, since we know $\mathbf{F}$, we have all the information we needed.
The function is more steep while $\beta$ is small, and is smooth while $\beta$ is relatively large.

However, if we put significant more data on the narrow steep region (e.g. $\beta \in (0,5)$), that may cause confusion on the network\textemdash the network will have bad performance on the wider smooth region since it does not see enough data. In order to achieve optimal overall performance, one need to balance between fitting the rough region and giving enough data of other regions.

First of all, we need a way to measure the ``roughness'' of the function in a  {given} area according to the parameter $\beta$. We choose how far away the thermal state $\rho = \exp(\sum_i\beta a_i F_i)/\tr[\exp(\sum_i\beta a_i F_i)]$ is from being a pure state as the indicator (denote as $\lambda$). In other words,
\[
\lambda = 1-\lambda^0
\]
where $\lambda^0$ is the biggest eigenvalue of $\rho$.

We divide $\beta$ into multiple intervals $I_i = (i,i+1]$ where $0 \le i \le 99$ and $i \in \mathbb{Z}$. In each interval $I_i$, 1,000 data points have been sampled. 1,000 $\beta$'s are drawn from uniform distribution in $I_i$ while 1,000 normalized $\vec a_i$ is sampled from normal distribution. These $\beta$'s, $\vec a$'s and $\mathbf{F}$ together form 1,000 $\rho = \exp(\sum_i\beta a_i F_i)/\tr[\exp(\sum_i\beta a_i F_i)]$. Getting $\lambda$ from each $\rho$, we calculate the average of these $\lambda$'s and denote it as $\bar\lambda_i$. The vector $\vec{\bar\lambda} = (\bar\lambda_1,\cdots,\bar\lambda_N)$ for all intervals is a characterization of the model according to the change of $\beta$   {(denote $N$ as the number of intervals for generality)}. Let $p_i = \bar\lambda_i/\sum_i \bar\lambda_i$ and
\[\vec{p}_{\beta} = (p_1,\cdots, p_N).\]
One may consider to use $\vec p_{\beta}$ to be the data distribution. But it transpired that $\vec p_{\beta}$ is not appropriate since it will concentrate the training data at the lower region.

Referring to our previous arguments, we need to balance the distribution. We take two flatten steps:

1) take the average of the first 10 elements in $\vec{\bar\lambda}$ and call it $\bar\lambda_\text{ten}$, then replace these first ten elements which are smaller than $\bar\lambda_\text{ten}$ with it;

2) denote $\sum_i \bar\lambda_i/N$ as $\bar\lambda_\text{avrg}$ and then replace elements which are smaller than $\bar\lambda_\text{avrg}$ with it.

We normalize the resulting vector and denote it as $\vec{p}_\text{flat}$. It is the data distribution we use in this work. Three different data generating methods have been compared in details in Appendix~\ref{sec:app_data}.


The neural networks used in this work are fully-connected feed-forward. It means the neurons in one layer is fully-connected to the neurons in the next layer and information only passes forward. The input and output layers are determined by the giving length of the measurement results (i.e. the cardinality of the fixed operator set $\mathbf{F}$). The three random 64 by 64 operator case have three input and three output neurons (we refer it as case 1 later in this paper). The 5-qubit 1D lattice example has 51 neurons ($5 \binom{3}{1}+4\binom{3}{1}\binom{3}{1}$ = 51) for input and output layers (We call it case 2). These two networks all have two hidden layers, each layer has 100 neurons.

The networks in this work are trained with Adam optimizer~\cite{kingma2014adam} which is a popular adaptive learning rate optimization algorithm designed for deep networks. The loss function we chose is Mean Absolute Error (MAE)
\[L(\vec e) = \frac{\sum_i^m |e_i|}{m}.\]
  {where $\vec e = \vec a - \vec a'$ is the error vector between the true value $\vec a$ and the estimated value $\vec a'$.}
MAE performs better than Mean Squared Error (MSE, $L(\vec e) = \sum_i^m e_i^2/m$, another commonly used loss function). It is because the parametrization function $P$~(\cref{eq:para_th}) would require $\vec a'$ to be as close to $\vec a$ as possible to bring the images close and
the square in MSE will make small errors more indistinct.

For case 1, the training data is 3,010,470. The batch size is 40,000 and we train the network for 300 epochs. And we use 2,005,584 pairs of training data for the 5-qubit 1D lattice model. The batch size is 20,000 and the number of epochs is also 300.

\subsection{Numerical results}\label{sec:res}




\begin{figure*}[t]
	\subcaptionbox{Case one: randomly generated operator set $\mathbf{F}$, 3 operators, dimension $d = 64$ \label{fig:rand64_uni}}
	{\includegraphics[width = .49\linewidth]{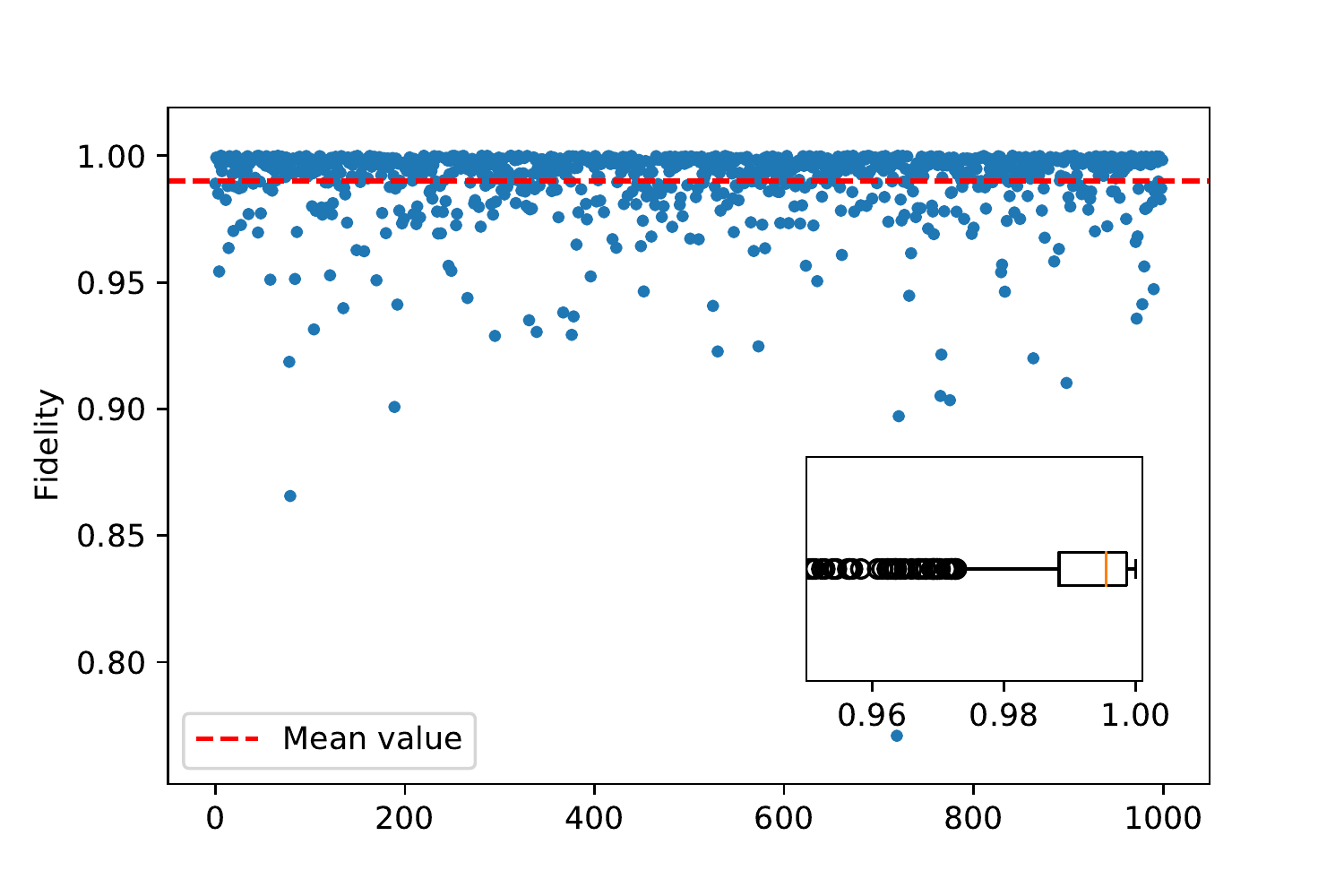}}
	\subcaptionbox{Case two: 5-qubit 1D lattice \label{fig:5qubit}}{\includegraphics[width = .49\linewidth]{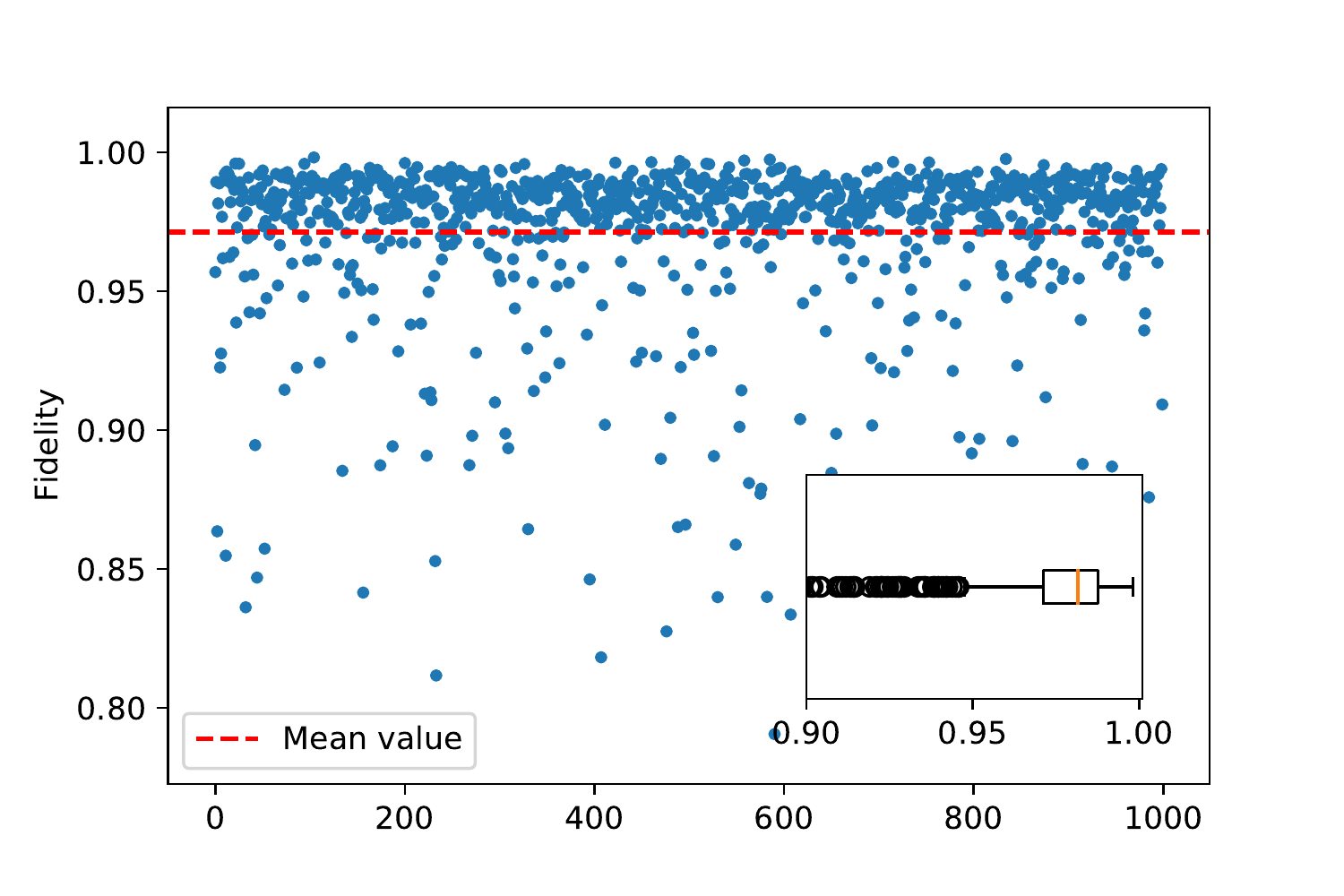}}
	\caption{\textbf{Testing results:} 1,000 pairs of new data has been tested for both cases. The mini-plots are the boxplots of these fidelities. The average fidelity for one is 99.0\%, for two is 97.1\%. }
\end{figure*}

New data sets are generated to test the performance of trained neural networks.
Similar to the procedure of producing training data in~\Cref{fig:data_gen}, the testing data are pairs of $\vec c$ and $\beta\vec a$.
$\beta$'s are uniformly picking from $(0,100]$ and $\vec a$'s are normalized.

The estimated MEE $\rho_\text{est}$ comes out from adopting the course in~\Cref{fig:network}. We compare each $\rho_\text{est}$ with its true MEE $\rho_\text{MEE}$ by calculating the fidelity.
The fidelity function we are using is the standard~\cite{nielsen2002quantum}
\[
f(\rho_1,\rho_2) = \tr\left(\sqrt{\sqrt{\rho_1}\rho_2 \sqrt{\rho_1}}\right).
\]

For case one, the average fidelity between true MEE $\rho_\text{MEE}$ and the estimated MEE $\rho_\text{est}$ is 99.0\%. \Cref{fig:rand64_uni} shows the fidelities of all tested data. The mini-figure is its boxplot~\cite{frigge1989some}, which is a graphical way to depict data through their quartiles. The orange line in the boxplot is the median value which is 99.5\%. Statistically, the circles in the boxplot are outliers which are data points notably different from others hence lack of statistical significance. Similarly, \Cref{fig:5qubit}  {shows} the fidelities of the whole testing data set for case two. The average fidelity is 97.1\% and median fidelity is 98.1\%.

\begin{table}
\begin{tabular}{ |c|c|c|c| }
\hline
 & Mean & Median & STD \\\hline
Case 1 & 99.0\% & 99.5\% & $17.0\times 10^{-3}$\\\hline
Case 2 & 97.1\% & 98.1\% & $31.1\times 10^{-3}$\\\hline
\end{tabular}
\caption{Statistics of numerical results}
\end{table}


\subsection{Experimental Verification and the effect of error}

\begin{figure*}[t]
\includegraphics[width=0.9\textwidth]{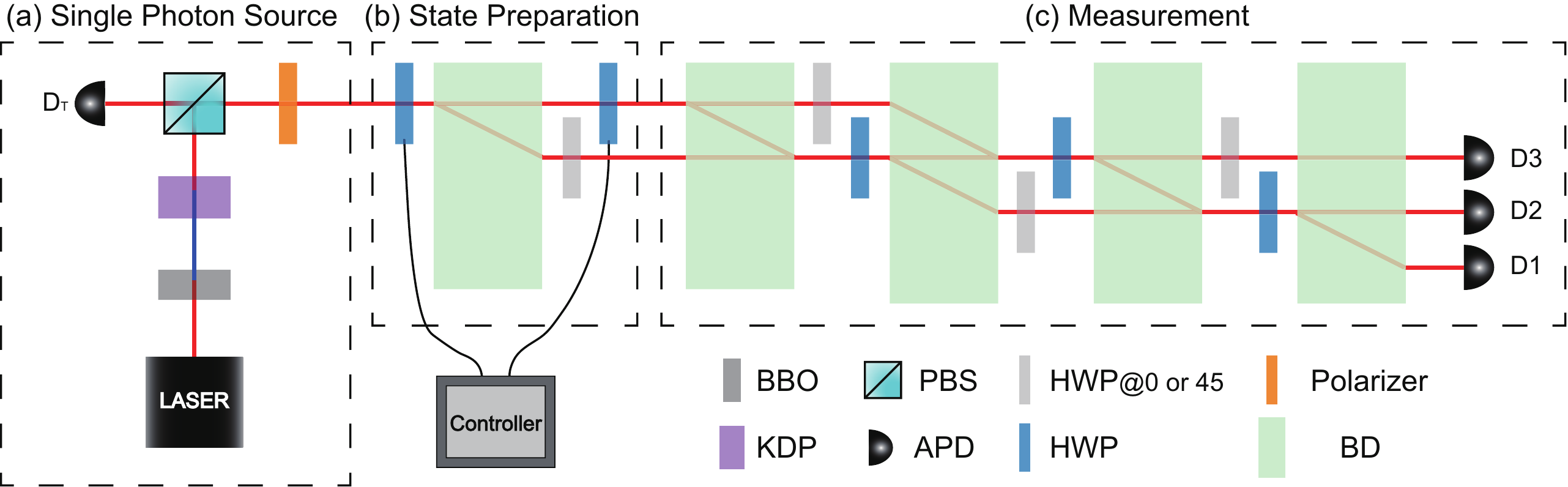}
\caption{Experimental set-up. (a) By pumping a phase-matched bulk potassium dihydrogen phosphate (KDP) crystal with the second harmonics generated from the beta barium borate ($\beta$-BBO) crystal, single photon pairs are generated. After a polarizing beam splitter (PBS), the idler mode is detected by the detector $\rm{D_T}$ as a trigger of the heralded single photon source, whereas the signal mode is directed towards the following set-up. (b) Through the half-wave plates (HWPs) and a calcite beam displacer (BD), single photons are prepared as photonic qutrits encoded in the polarization and spatial modes. (c) The measurement part is composed of wave plates and BDs which form a three-stage interferometer capable of implementing arbitrary qutrit unitary. After the unitary transformation, photons are detected by three avalanche photodiodes (APDs). By setting the wave-plates with different angles, measurement of different operators can be realized.}\label{fig:exp_setup}
\end{figure*}

To verify the performance of our well-trained neural network in processing real experimental data and its robustness against experimental noise, we implement a qutrit photonic set-up capable of preparing different qutrit states and measuring arbitrary operators, as shown in~\Cref{fig:exp_setup}.
Particularly, when experimental data are generated by unique ground states of a pseudo Hamiltonian, $\rho_\text{MEE}$ is the exact ground state been measured, and the NN estimation $\rho_\text{est}$ is also the approximation of the real state.
Therefore, we intentionally prepare ground states of a class of pseudo Hamiltonians and feed them into the measurement devices. By directly comparing the prepared ground states $\rho_\text{exp}$ with $\rho_\text{est}$, we can reveal the network's performance in real experiments.

In our experiment, we choose two set of operators $\mathbf{F}_1$ and $\mathbf{F}_2$, and each contain 3 hermitian operators (see explicit expression in Appendix~\ref{sec:app_exp}). For each set, 300 ground states $\{\rho_{\rm{exp}}\}$ of the pseudo Hamiltonian are randomly prepared by changing the setting angles of the two configurable half-wave plates (HWPs) in~\Cref{fig:exp_setup} (b) (see Appendix~\ref{sec:app_exp} for details). Then the prepared states are input into the measurement part, which is constituted by wave plates, calcite beam displacers (BDs) and photon detectors, capable of projecting the input states into an arbitrary basis. From the measurement statistics, expectation values of different operators can be estimated (see Appendix~\ref{sec:app_exp} for details). Thus by this preparation-and-measurement set-up, we obtain the experimental data set $\vec c_\text{exp} =( c_{1,\text{exp}},c_{2,\text{exp}},c_{3,\text{exp}} ) $.

\begin{figure*}[t]
\begin{subfigure}{.5\textwidth}
  \centering
  \includegraphics[width=\linewidth]{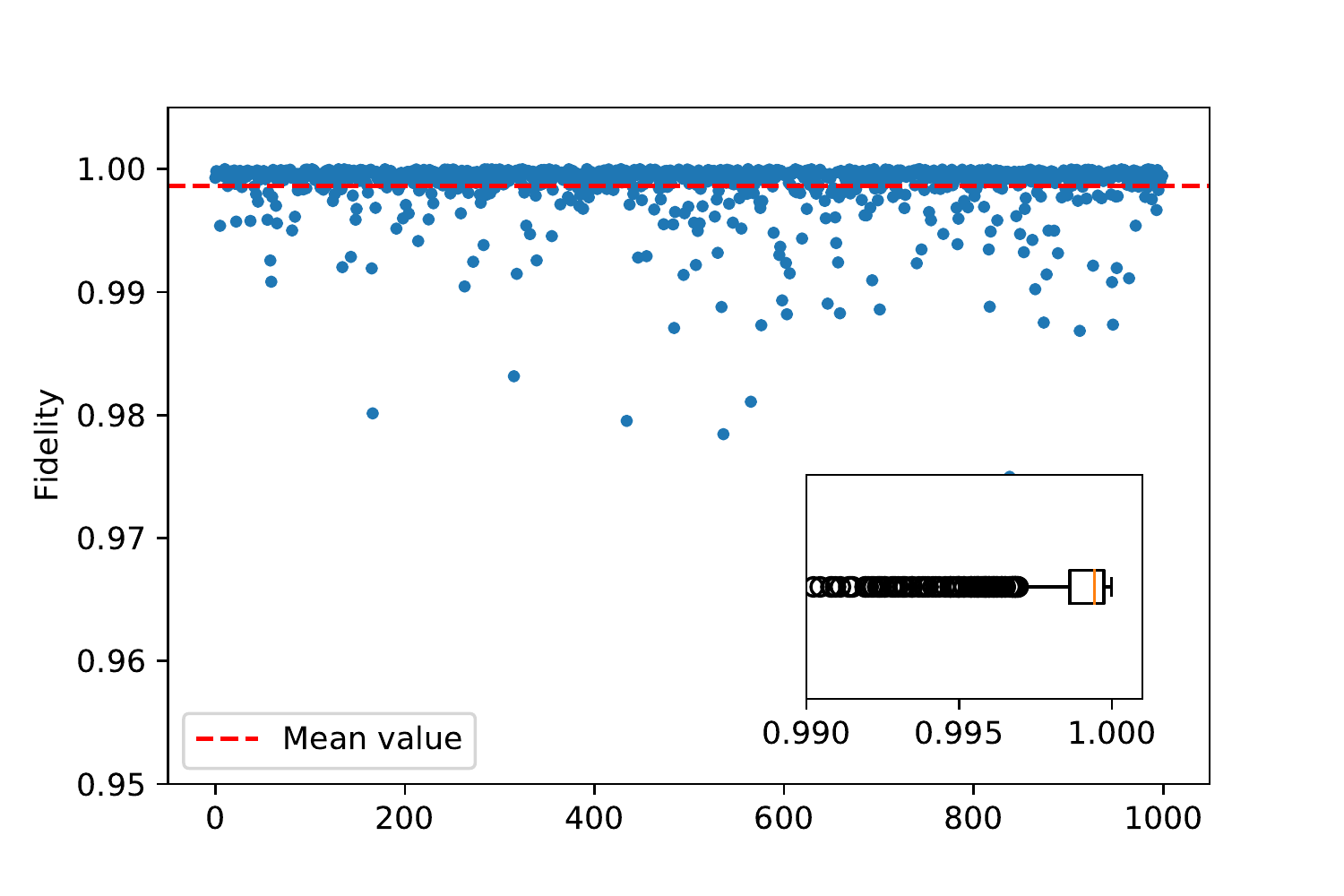}
  \caption{\textbf{Numerical testing results of $\mathbf{F}_1$:} Blue dots are the fidelities between the true MEE state $\rho_\text{MEE}$ and the estimated state $\rho_\text{est}$. The x-axis is the dummy variable of the testing set. 1,000 data has been tested, the mean value is 99.9\%}\label{fig:new3_num}
\end{subfigure}
\begin{subfigure}{.48\textwidth}
  \centering
  \includegraphics[width=\linewidth]{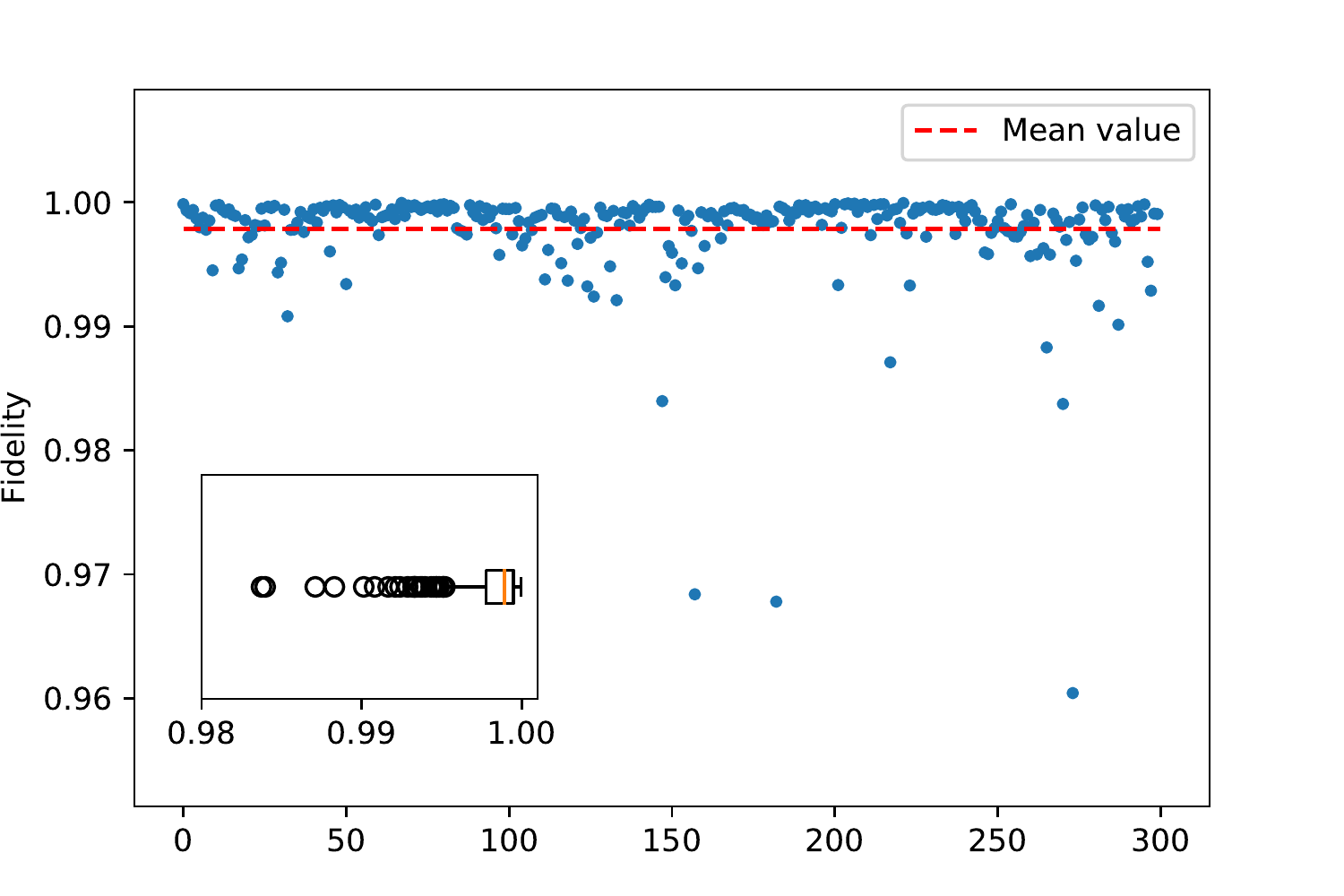}
  \caption{\textbf{Fidelities with experimental data of $\mathbf{F}_1$:} The horizontal axis is the dummy label of experimental data points. The average fidelity for all 300 data points is 99.8\%. The median value is 99.9\% (orange line in the boxplot).}
  \label{fig:new3_exp}
\end{subfigure}

\begin{subfigure}{.49\textwidth}
  \centering
  \includegraphics[width=\linewidth]{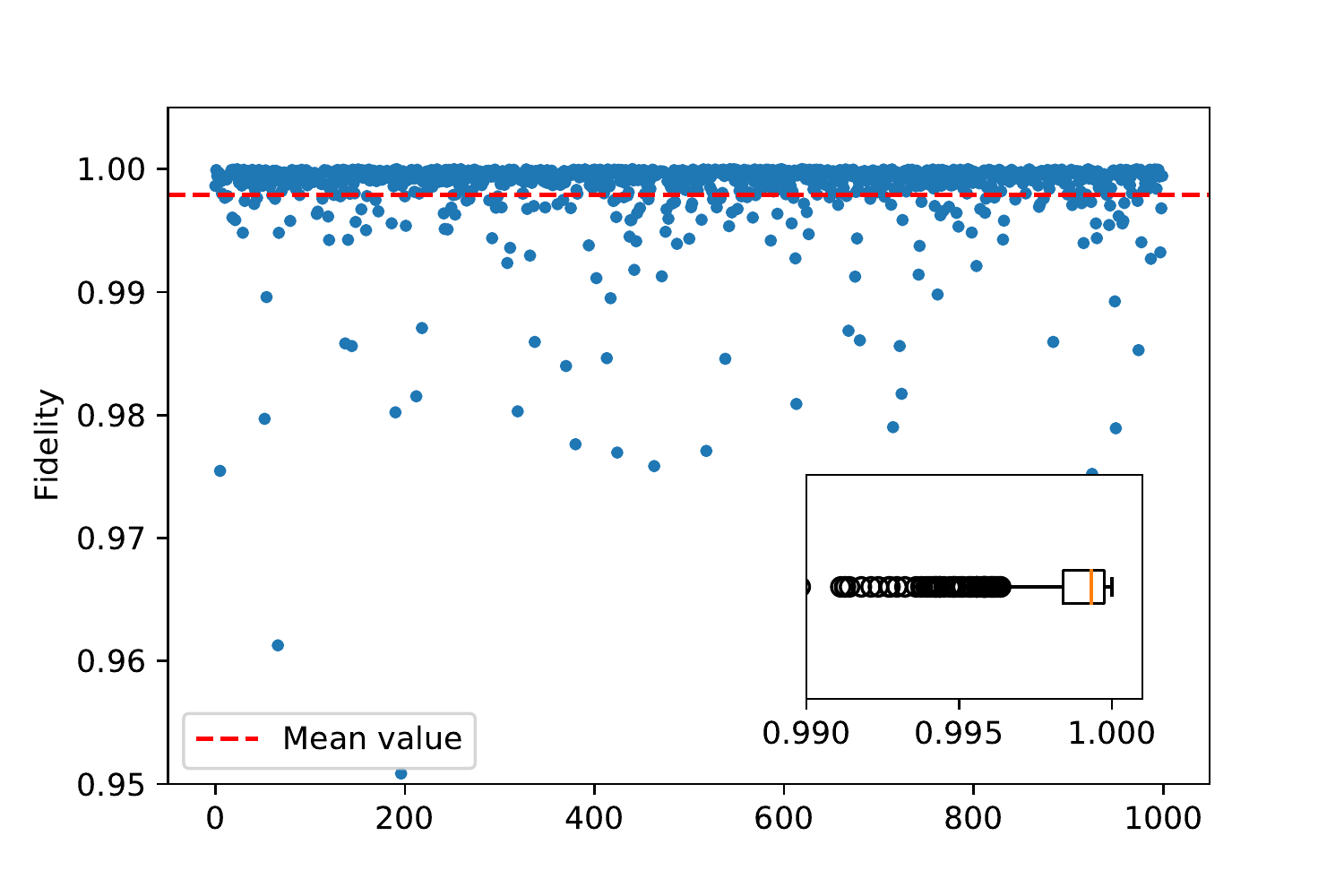}
  \caption{\textbf{Numerical testing results of $\mathbf{F}_2$:}Blue dots are the fidelities between the true MEE state $\rho_\text{MEE}$ and the estimated state $\rho_\text{est}$. The x-axis is the dummy variable of the testing set. 1,000 data has been tested, the mean value is 99.8\%}
  \label{fig:new5_num}
\end{subfigure}
\begin{subfigure}{.49\textwidth}
  \centering
  \includegraphics[width=\linewidth]{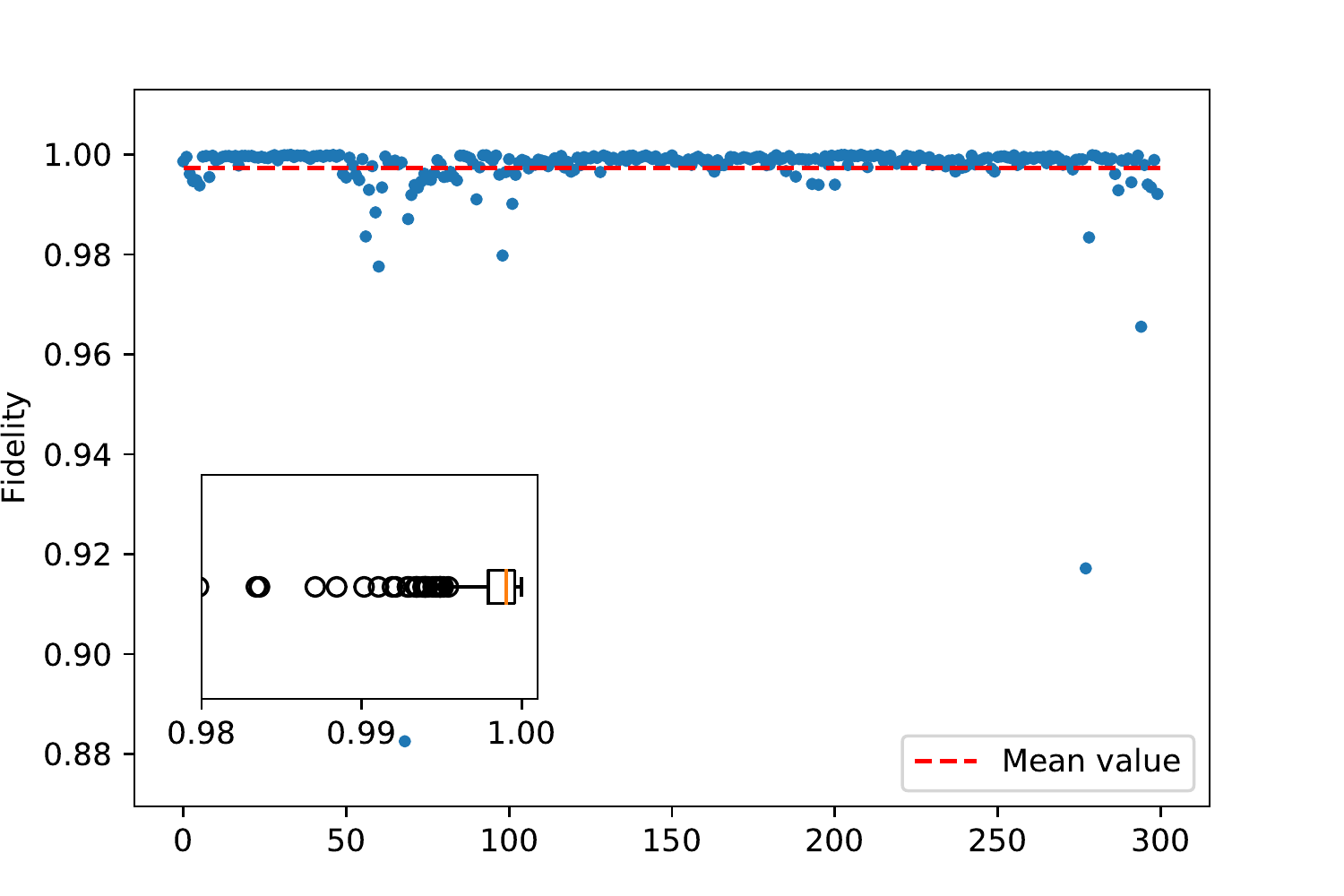}
  \caption{\textbf{Fidelities with experimental data of $\mathbf{F}_2$:} The average fidelity of 300 data points is 99.7\%. The median value shows in the boxplot is 99.9\%. }
  \label{fig:new5_exp}
\end{subfigure}
\caption{Numerical and experimental fidelities of $\mathbf{F}_1$ and $\mathbf{F}_2$}
\label{fig:exp_result}
\end{figure*}

Before feeding experimental data into the neural networks, we have trained the networks individually for each operator set. 1,010,196 and 1,003,808 pairs of numerical data have been used to train networks for $\mathbf{F}_1$ and $\mathbf{F}_2$, respectively. The network structure and other settings (e.g. training algorithm, loss function etc.) are in similar fashion with the previous numerical cases. \Cref{fig:new3_num} shows the numerical results of $\mathbf{F}_1$ for 1,000 random generated data. The average fidelity is 99.9\%. \Cref{fig:new5_num} is the testing fidelities for $\mathbf{F}_2$, the mean value is 99.8\%.

The well-tuned neural networks are now ready for experimental data.
Measurement outcomes $\vec c_\text{exp}$ derived from the experiments are inputs of the networks.
From the output parameter set $\beta' \vec a'$, the estimated MEEs $\rho_{\rm{est}}$'s can be derived. The fidelities between $\rho_{\rm{exp}}$ and $\rho_{\rm{est}}$ have been calculated and are shown in \Cref{fig:new3_exp} ($\mathbf{F}_1$) and \Cref{fig:new5_exp} ($\mathbf{F}_2$). The mean value of all 300 data points is 99.8\% for $\mathbf{F}_1$, and is 99.7\% for $\mathbf{F}_2$.

In this experiment, the measurement outcomes are suffered from different systematic errors, such as inaccuracies of wave plate setting angles, imperfect interference visibility and drifting of the interferometers, and statistical fluctuations. The average relative errors of different operators ($(c_{i,\text{exp}}-\tr(\rho_{\rm{exp}} F_i)/{\tr(\rho_{\rm{exp}} F_i})$) range in $0.79\%\sim2.43\%$ (see more details in Appendix~\ref{sec:app_exp}). Even in this level of additional experimental errors, the networks show almost the same performance in processing experimental data compared with numerical data.


\subsection{In comparison with other methods}

The maximum entropy estimation
\[
\rho_\text{MEE} = \frac{\exp(\beta\sum_i a_i F_i)}{\tr[\exp(\beta\sum_i a_i F_i)]},
\]
for given $\vec c = (\tr(\rho F_1),\cdots,\tr(\rho F_m))$ is an optimization problem. It is closely related
to the field of information geometry, statistical inference, and machine learning.

An iterative algorithm based on information geometry viewpoint is propose in~\cite{niekamp2013computing},  which runs as follows. First, initialize the system Hamiltonian as an identity operator $H = I$, so the initial density matrix $\rho_{\text{ini}}=\exp(I)/\tr \exp(I)$ is the maximum mixed state. The following task is to solve the equations $\tr (\rho F_i)=\tr (\tau F_i)$ for each $i$, or, to be more precisely, find a density matrix $\tau$ to minimize $\sum_i |\tr (\rho F_i)-\tr (\tau F_i)|$. This is done by iteratively update the Hamiltonian $H$ by $H+\epsilon F_i$, so that the density matrix $\tau$ is updated as
\[
\tau=\frac{e^{H}}{\tr e^{H}}\rightarrow \tau'=\frac{e^{H+\epsilon F_i}}{\tr e^{H+\epsilon F_i}},
\]
in which, the parameter $\epsilon$ is something like a gradient and could be approximated as		
\[
\epsilon=\frac{\tr F_i \rho-\tr F_i\tau}{\tr F_i^2\tau-(\tr F_i\tau)^2}
\]
for each $F_i$. Repeat the iteration for several times and we can find a $\tau$ as closely to $\rho$ as possible.


Another related method is base on the so-called quantum Boltzmann machine (QBM)~\cite{amin2018quantum}.
The QBM uses a different loss function (or objective function) for optimization, i.e. the cross entropy,
\[
\mathcal{L}=-\sum_i p_i \log p_i',
\]
with $p_i$ and $p_i'$ are probability distributions: $p_i$ is the ideal case and $p_i'$ is relative to some parameters. The learning process of a quantum Boltzmann machine is to find certain parameters to minimize $\mathcal{L}$. Take $p_i=C\tr \rho F_i$ and $p_i'=C'\tr \tau F_i$, where $C$ and $C_i'$ is a normalization constant. The densit matrix $\tau$ here could also be expressed as $\tau=\exp(H)/\tr \exp(H)$. Since $H=\sum_i a_i F_i$, the loss function is now a function of $a_i$s. The loss function $\mathcal{L}$ reaches its minimum for $p_i=p_i'$, so our goal is to optimize $\mathcal{L}$ over possible $a_i$s.

We can use the same method which  the QBM use to learn the maximum entropy state.
To use the cross entropy, for $F_i$s with negative eigenvalues, we first renormalize $p_i'$s by adding $(\lfloor-f_{i\text{min}}\rfloor+1)I$ to $F_i$, where $f_{i\text{min}}$ is the lowest eigenvalue of $F_i$. This ensures $p_i'$s being positive, and adding unity operator to Hamiltonian has no effect on its thermal state. Second, the $p_i$ and $p_i'$ in cross entropy are probability distributions, which means $\sum_i p_i$ and $\sum_i p_i'$ are both restricted to $1$, so we add normalization constants $C$ and $C'$ in front of $\tr \rho F_i$ and $\tr \tau F_i$, respectively.

We test both the iterative algorithm and the QMB algorithm using MATLAB, for the examples in Appendix B. The iterative algorithm converges to the desired results precisely and effectively. The average time for an iterative algorithm for each case is about 0.0425 seconds.
As a comparison, if we run the optimization using the functions provided by MATLAB, the time for each case is about 0.0148 seconds.

The method of QBM, however, cannot provide a precise approximation to the original density matrix.
This may due to the fact that the gradient is hard to obtain, (notice that the forms of matrix $F_i$ in our cases are far more complicated than that the ones discussed in QBM (see~\cite{amin2018quantum}). Also, it could due to the normalization of $p_i$s we have introduced, would introduce more troubles in the learning process. There could be ways to improve the training method, which we will leave for future investigation.

Given that the iterative algorithm seems more effective and accurate for
optimization, we will then compare our supervised learning method with the iterative algorithm.
For the case that the measured set $\mathbf{F}$ possess three 64 by 64 hermitian operators, our method estimates the test set with 99.0\% average fidelity~(Section~\ref{sec:res}). Setting the error bound as $10^{-10}$. As a comparison, the iterative algorithm provides the outcome states with fidelity almost $1$ for every data point. In terms of accuracy, the interactive algorithm is slightly stronger than ours.

By using the same computational device~\footnote{Macbook Pro, Processor 2.3GHz Quad-Core Intel Core i5, 8GB Memory}, our network could predict 5,000 data in less than a second while the iterative method requires about 10 minutes for 100 data. In this sense, once trained, our method is more efficient for estimation without loss of too much accuracy.

\section{Discussion}


In this paper, we presented a deep learning method for Quantum Inverse Problems. The method shown good performances for both numerical and experiment data, and robustness against experimental noise.

The example we demonstrated is a quantum state learning problem, the traning data is numerically generated. The network can also train with noisy experimental data to output idea (noiseless) state. The outcomes from the trained network would naturally mitigate the experimental error.
Our method can also easily adapt to other setups such as Hamiltonian learning. An example and detailed discussion can be found in~\cite{cao2020supervised}.

We show that our method is straightforward to implement for scenarios that prior information grantees a most bijective relation between $\rho$ and $\vec c$. It can also use to exam if the prior information is actually adequate to ensure the map to be almost bijective. After properly implemented this scheme, if the trained network does not perform well. We may consider that the prior information is not exactly pined down a almost one-to-one correspondence.

For more ill-posed problems, there are several techniques can be applied, such as various regularizers~\cite{li2020nett}, different network structures~\cite{jin2017deep}, statistical estimators~\cite{adler2018deep,adler2018learned}, depending on the particular problem.

The last but not least, the neural networks in the inverse process can be replaced by quantum neural networks~\cite{farhi2018classification}. In this case, we may not need the parametrization process $P$. The modified method has a potential to implement on NISQ devices.


\begin{acknowledgments}
N.C. acknowledges the Natural Sciences and Engineering Research Council of Canada (NSERC).
J.X., A.Z. and L.Z. were supported by the National Key Research and Development Program of China (Grant Nos. 2017YFA0303703 and 2019YFA0308704) and the National Natural Science Foundation of China (Grant Nos. 91836303, 61975077, and 11690032).
\end{acknowledgments}


\appendix
\section{Influences of training data distribution}\label{sec:app_data}

In this section, we show the influence of three different training $\beta$ distribution on the neural network performance: 1) evenly distributed $\vec p_\text{even} = (1/N, \cdots, 1/N)$; 2) the distribution $\vec p_{\beta}$ mentioned in the main text which only considered the roughness of $\beta$; 3) and the flattened distribution $\vec p_\text{flat}$ that we used in this work. (The technical  definitions see Section~\ref{sec:data_pre}.)

\begin{figure}[ht]
  \centering
  \includegraphics[width=\linewidth]{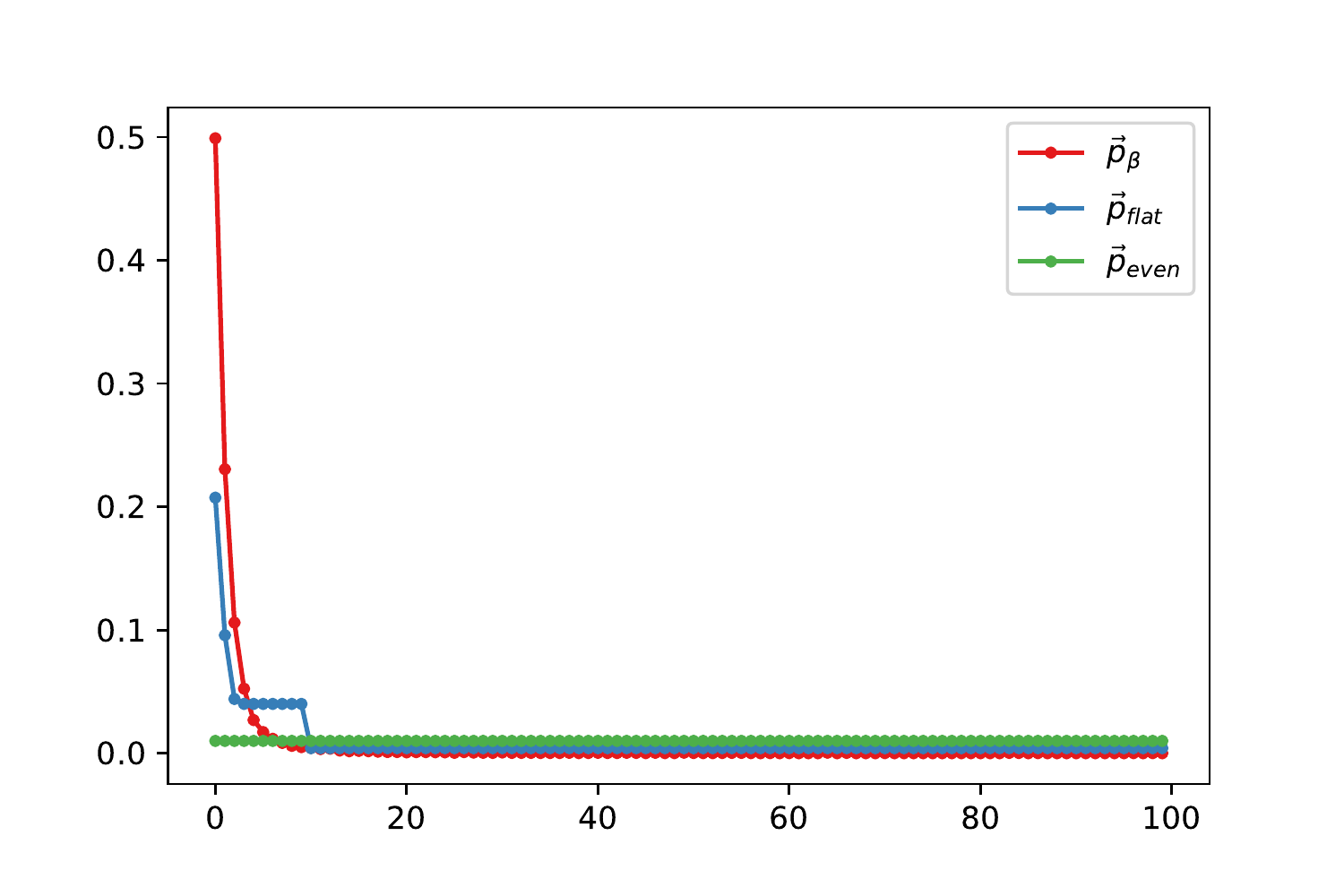}
  \caption{The three distributions for the operator set $\mathbf{F}_1$: The x-axis is the label $i$ of the interval $I_i = (i,i+1]$ for $\beta$. The green, blue, red dots depict the even distribution $\vec{p}_\text{even}$, the flattened distribution $\vec p_\text{flat}$ and $\vec p_{\beta}$ respectively.}
  \label{fig:n3_dist}
\end{figure}

We consider the operator set $\mathbf{F}_1$ in Appendix~\ref{sec:app_exp}. The three distributions for $\mathbf{F}_1$ are shown in~\Cref{fig:n3_dist}.
The horizontal axis is the index $i$ of interval $I_i = (i,i+1]$. The vertical axis shows the percentage of how much $\beta$'s are sampled from a giving interval $I_i$. $\vec{p}_{\beta}$ is dominantly concentrated on the first few intervals.
We train three networks separately with each distribution.
To fairly compare them, we prepare the same amount of training data for each one, and using exactly same training settings.
The number of training data is about 1,000,000 (round up the number when the distribution multiply by 1,000,00 does not get integers). 

\begin{figure*}
\begin{subfigure}{.49\linewidth}
  \centering
  \includegraphics[width=\linewidth]{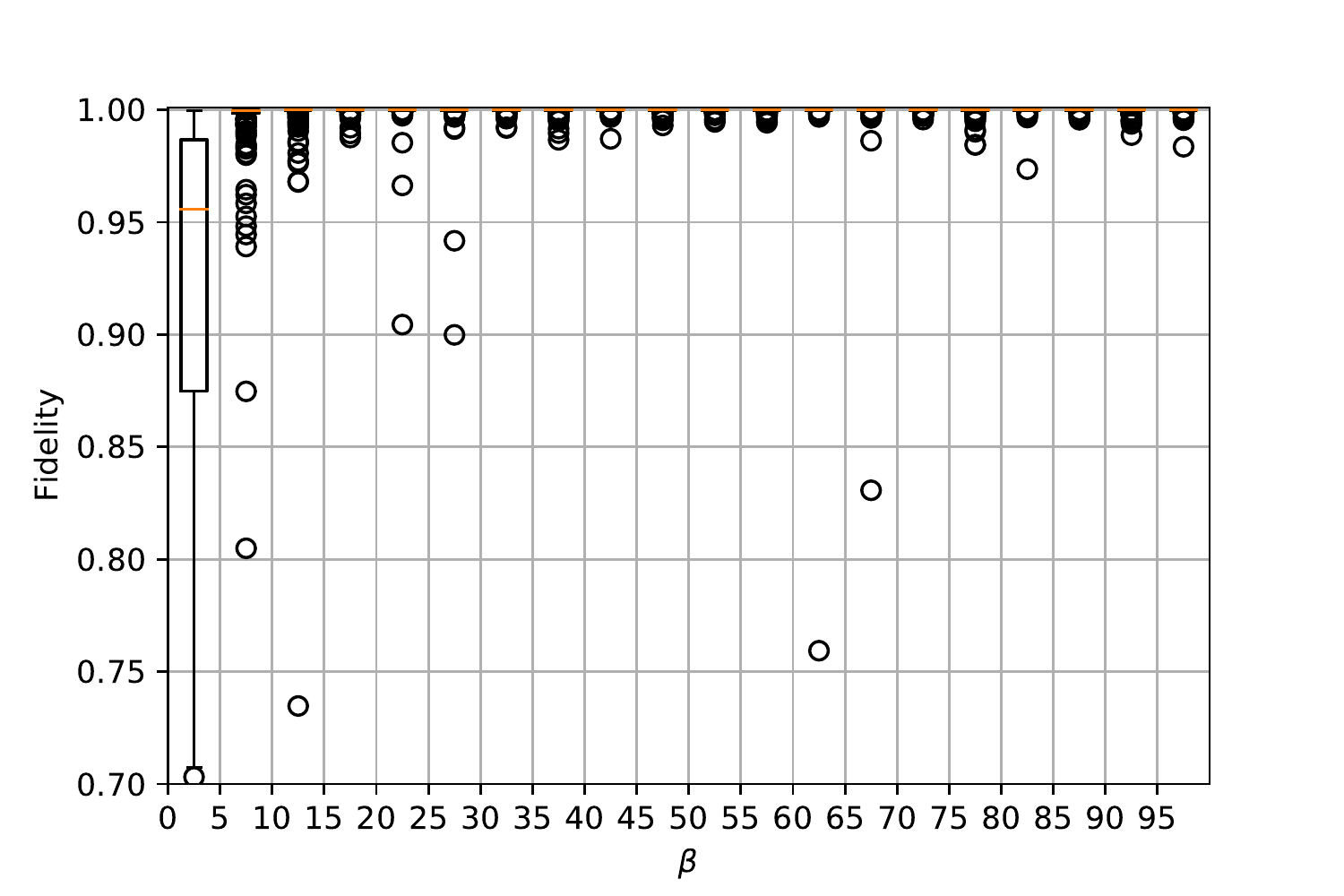}
  \caption{\textbf{Boxplots of $\vec{p}_\text{even}$:} every boxplot represents the fidelities of 5 intervals $I_i$. The even distribution $\vec{p}_\text{even}$ has bad performance on the first 5 intervals.}
  \label{fig:even_box}
\end{subfigure}
\begin{subfigure}{.49\linewidth}
  \centering
  \includegraphics[width=\linewidth]{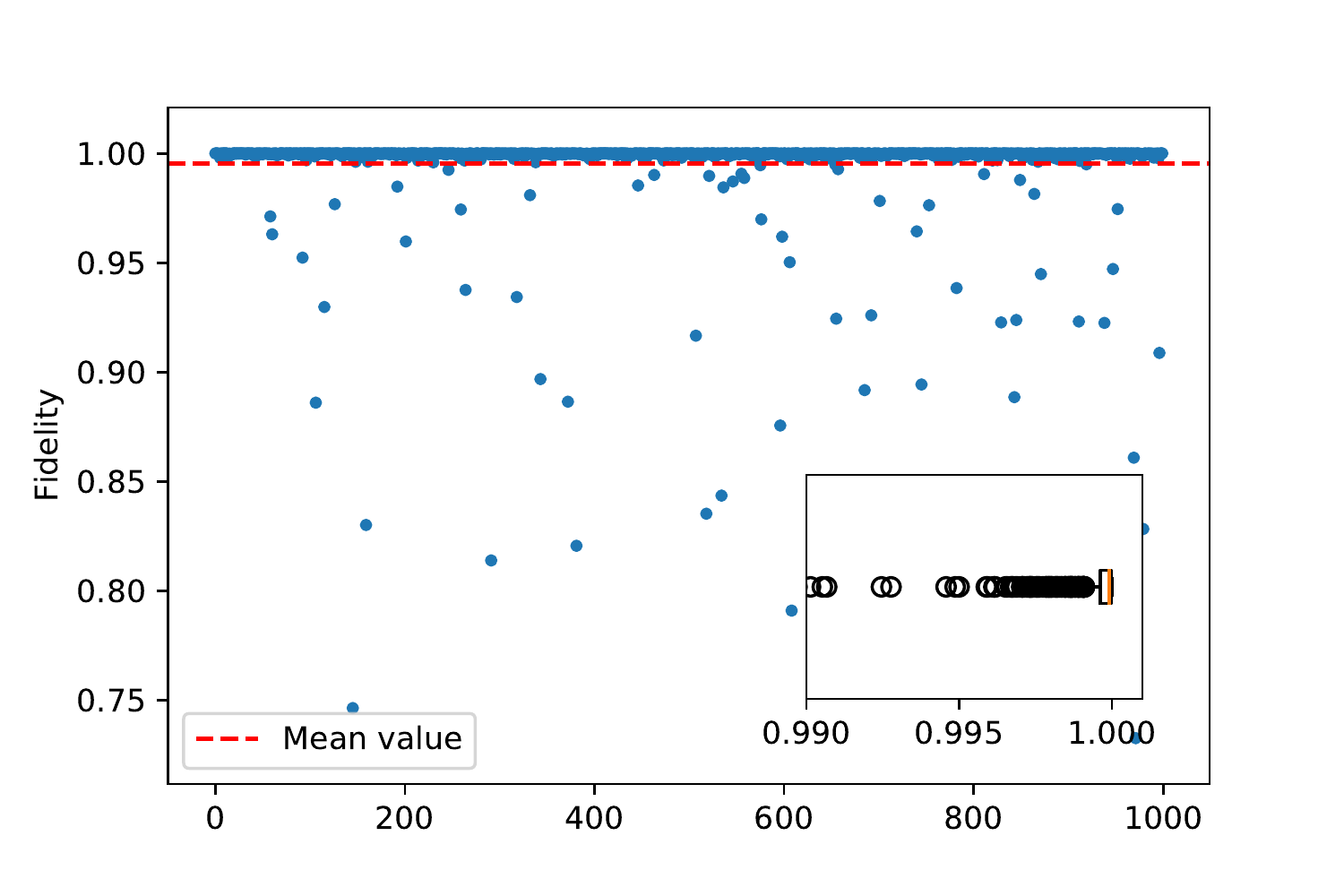}
  \caption{\textbf{Fidelities of the evenly generated test set ($\vec{p}_\text{even}$):} there are several points have extreme low fidelities, mostly comes form the low $\beta$ value region. The average fidelity is 99.5\% for 1,000 data.}
  \label{fig:even_uni}
\end{subfigure}
\begin{subfigure}{.49\linewidth}
  \centering
  \includegraphics[width=\linewidth]{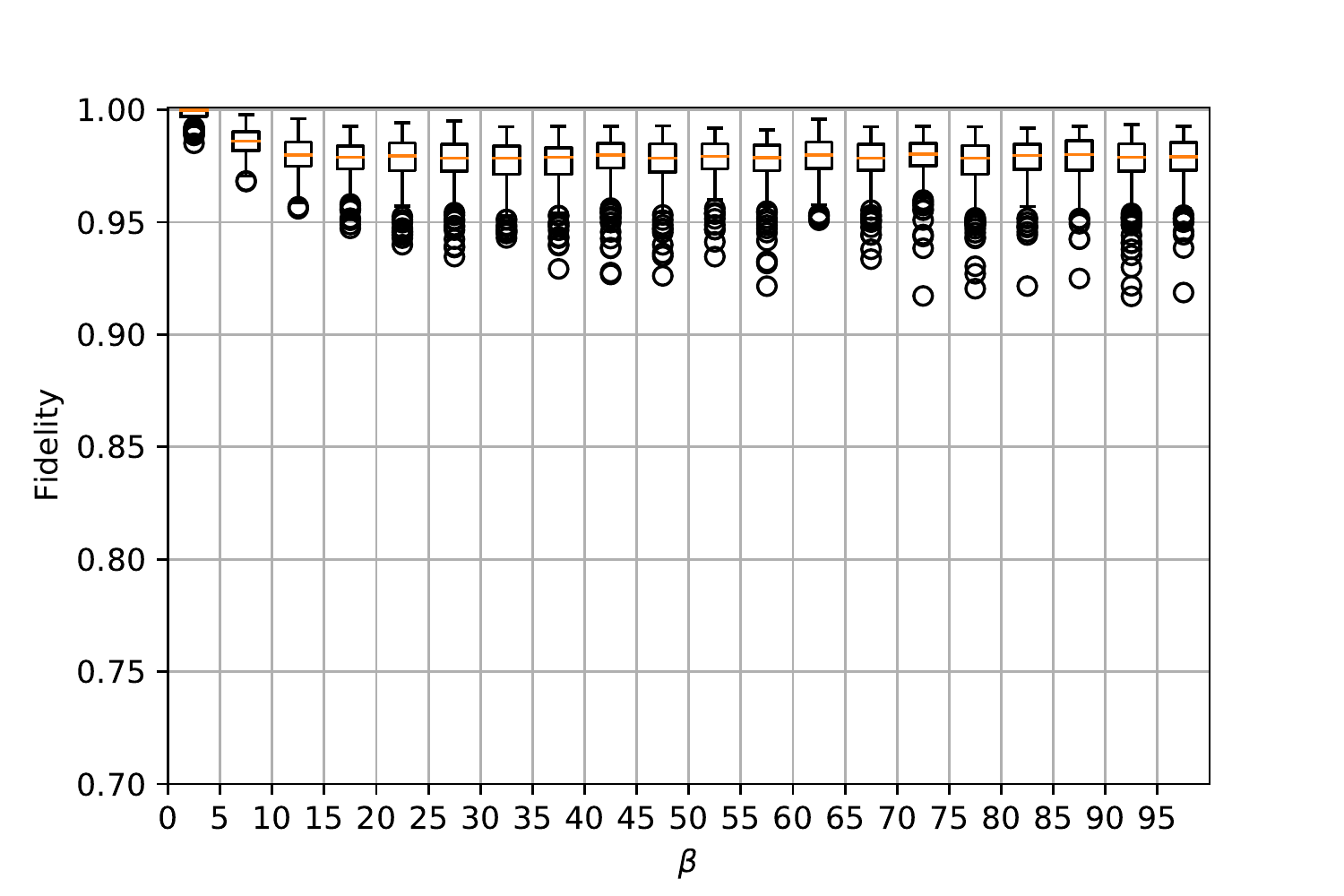}
  \caption{\textbf{Boxplots of $\vec{p}_\beta$:} every boxplot represents the fidelities of 5 intervals $I_i$. Except for the first five, the beta distribution $\vec{p}_\beta$ has substandard performance on most of the intervals.}
  \label{fig:beta_box}
\end{subfigure}
\begin{subfigure}{.49\linewidth}
  \centering
  \includegraphics[width=\linewidth]{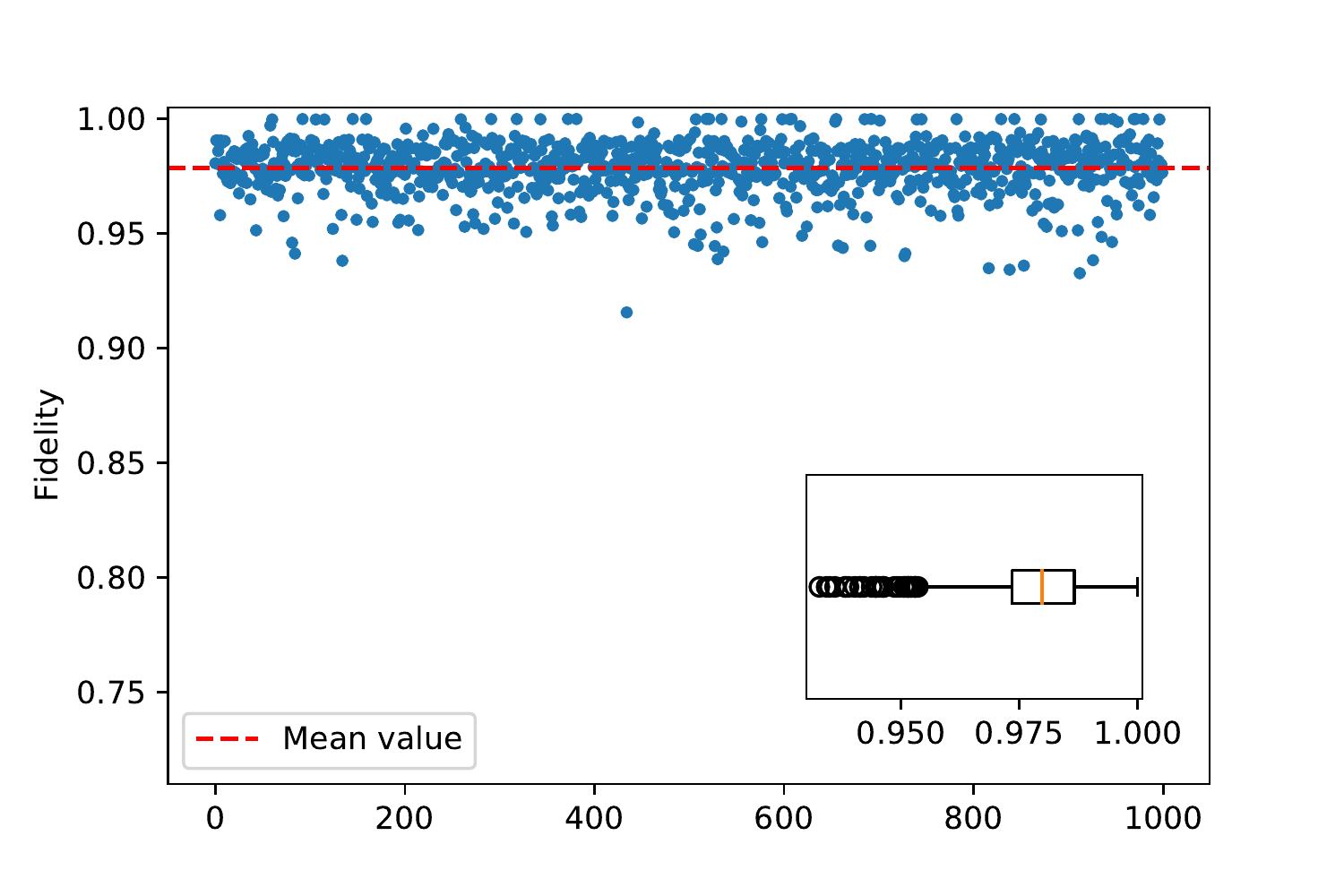}
  \caption{\textbf{Fidelities of the evenly generated test set ($\vec{p}_\beta$):} The standard deviation of the 1,000 tested results is significantly larger than the other two. The average is 97.9\%. }
  \label{fig:beta_uni}
\end{subfigure}
\begin{subfigure}{.49\linewidth}
  \centering
  \includegraphics[width=\linewidth]{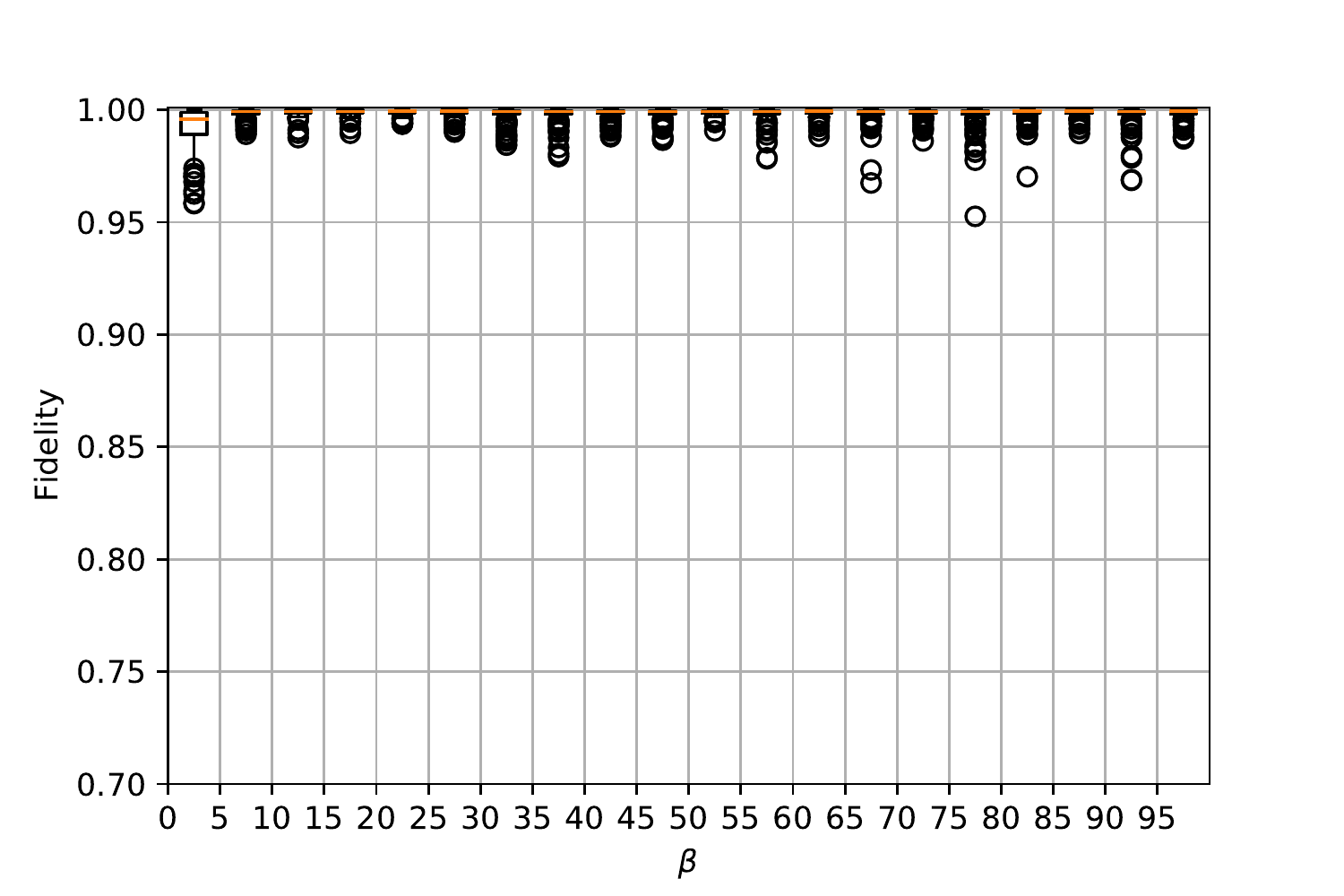}
  \caption{\textbf{Boxplots of $\vec{p}_\text{flat}$:} every boxplot represents the fidelities of 5 intervals $I_i$. Comparing to the other two distribution, the flattened distribution has the best overall performance.}
  \label{fig:flat_box}
\end{subfigure}
\begin{subfigure}{.49\linewidth}
  \centering
  \includegraphics[width=\linewidth]{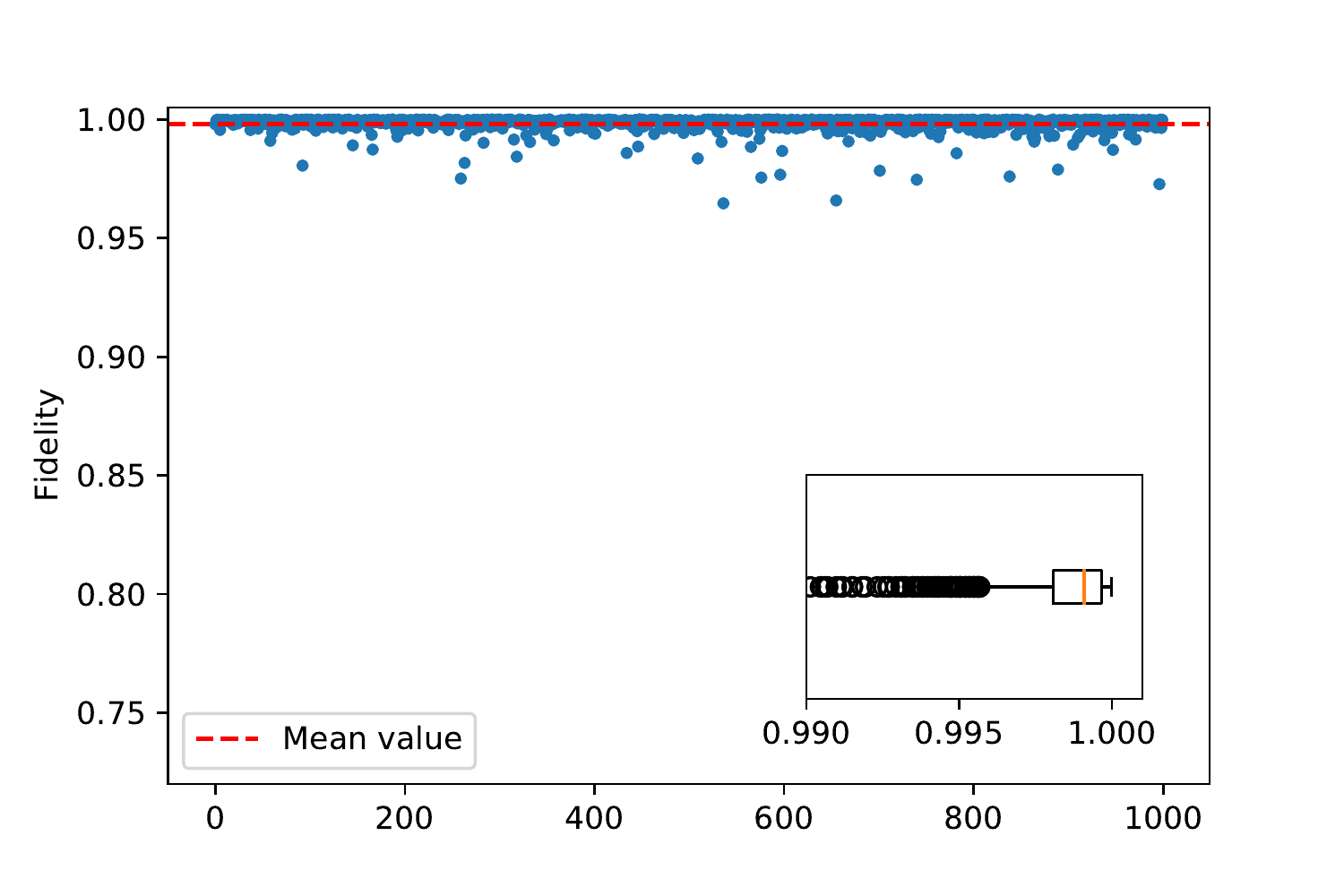}
  \caption{\textbf{Fidelities of the evenly generated test set ($\vec{p}_\text{flat}$):} The average fidelity is 99.9\%.}
  \label{fig:flat_uni}
\end{subfigure}
\caption{Results of two test sets for $\vec{p}_\text{even}$, $\vec{p}_\beta$ and $\vec{p}_\text{flat}$ (for comparison purpose, we use the same scale for each plot).}
\end{figure*}

Two testing data sets have been generated. Set one has 5,000 data points\textendash 50 different $\beta$'s have been uniformly drawn from every interval $I_i$. Fidelity boxplots of every 5 intervals present in~\Cref{fig:even_box} ($\vec{p}_\text{even}$),~\Cref{fig:beta_box} ($\vec{p}_{\beta}$) and~\Cref{fig:flat_box} ($\vec{p}_\text{flat}$).
For comparison purpose, we use the same scale for each plot.
The network tuned with the even distribution $\vec{p}_\text{even}$ data set has significantly poor performance when $\beta\in (0,5]$ and also has several exceptional outliers on other intervals (\Cref{fig:even_box}). The network of $\vec{p}_\beta$ is expected to have high fidelity for $\beta\in (0,5]$ and substandard performance on other parts (\Cref{fig:beta_box}) because of the data concentration. The network of $\vec{p}_\text{flat}$ has a balance in between ((\Cref{fig:flat_box})).

The second test set has 1,000 data points which $\beta$'s are uniformly taken from $(0,100]$. The testing results are shown in~\Cref{fig:even_uni}($\vec{p}_\text{even}$),~\Cref{fig:beta_uni} ($\vec{p}_{\beta}$) and~\Cref{fig:flat_uni} ($\vec{p}_\text{flat}$).

\section{Experimental details}\label{sec:app_exp}

The two qutrit operator sets $\mathbf{F}$ in our experiment are as follows $\mathbf{F}_1 = \{F_{11}, F_{12}, F_{13}\}$, $\mathbf{F}_2 = \{F_{21}, F_{22}, F_{23}\}$ where
\begin{eqnarray*}
      F_{11}=\begin{pmatrix} 0 & 1 & 0 \\ 1 & 0 & 0 \\ 0 & 0 & 0\end{pmatrix},
      F_{12}=\begin{pmatrix} 0 & 0 & 1 \\ 0 & 0 & 0 \\ 1 & 0 & 0\end{pmatrix},
      F_{13}=\begin{pmatrix} 1 & 0 & 0 \\ 0 & -1 & 0 \\ 0 & 0 & 0\end{pmatrix},\\
      F_{21}=\begin{pmatrix} 2 & 0 & 0 \\ 0 & 0 & 1 \\ 0 & 1 & 0\end{pmatrix},
      F_{22}=\begin{pmatrix} 0 & 0 & 1 \\ 0 & 0 & 0 \\ 1 & 0 & 0\end{pmatrix},
      F_{23}=\begin{pmatrix} 0 & 0 & 0 \\ 0 & 0 & 0 \\ 0 & 0 & 2\end{pmatrix}, \nonumber
\end{eqnarray*}
here $F_{ji}$ stands for the $i$th operator in the $j$th set. To demonstrate the neural network's performance, we sample 300 ground-states $\{\ket{\psi_{ji}}\}$ of pseudo Hamiltonian $\{H_{j}=\sum_{i=1}^3 a_{ji}F_{ji}\}$ by randomly ranging the parameter set $\{a_{ji}\}$. As shown in~\Cref{fig:exp_setup}, wave plates and a BD are use to distribute single photons in the superposition of optical polarization and spatial modes, realizing the preparation of these ground states. Note that only two configurable HWPs are enough for the preparation (no need for quarter-wave plates or other phase retarders), as the operator sets are all real operators and the ground states should also be real. The three eigen-modes of the qutrit state are defined as $\ket{0}=\ket{H}\otimes\ket{s1},\ket{1}=\ket{H}\otimes\ket{s2},\ket{2}=\ket{V}\otimes\ket{s2}$, where $\ket{H}$ ($\ket{V}$) stands for the horizontal (vertical) polarization and $\ket{s1}$ ($\ket{s2}$) stands for the upper (lower) spatial mode.

As for the measurement of different operators $F_{ji}$, we use linear optical devices such as wave plates and BDs to construct a three stage interferometer which is capable of implementing arbitrary qutrit unitary operation \cite{PhysRevLett.125.150401}. For the same reason, here only HWPs are needed and the set-up is relatively simpler than implementing an universal unitary. To estimate $\tr(\rho F_{ji})$, we apply the unitary transformation
\begin{equation*}\label{eqn:Unitary}
  U_{ji}=\ket{0}\bra{\lambda^{(ji)}_0}+\ket{1}\bra{\lambda^{(ji)}_1}+\ket{2}\bra{\lambda^{(ji)}_2}
\end{equation*}
on an input state $\rho$, here $\ket{\lambda^{(ji)}_k}(k=0,1,2)$ is the corresponding eigen-vector of $F_{ji}$ with eigen-value $\lambda^{(ji)}_k$. It transforms any state from the eigen-basis of $F_{ji}$ into computational or experimental basis. Therefore, from the measurement statistics measured by the following detectors, the expectation value $\tr(\rho F_{ji})$ of $F_{ji}$ can be estimated.

Throughout this experiment, the experimental errors are mainly contributed from systematic errors, as the statistical fluctuations are very low due to enough trials (above 35000 registered photons) for each measurement. The systematic errors include inaccuracies of wave plate setting angles (typically $\sim0.2$ degree) in state preparation and measurement stage and imperfections of the interferometers. Especially during the measuring progress, slow drift and slight vibrating of the interferometers will cause a decrease of the interference visibility. In our experiment, the interference visibilities are maintained above $98.5\%$. The average relative errors $(c_{i,\text{exp}}-\tr(\rho_{\rm{exp}} F_i)/{\tr(\rho_{\rm{exp}} F_i})$ of measured expectation values of different operators are shown in \Cref{tab:relative error}.

\begin{table}[ht]
\begin{tabular}{ |c|c|c|c| }
\hline
$(c_{i,\text{exp}}-\tr(\rho_{\rm{exp}} F_i)/{\tr(\rho_{\rm{exp}} F_i})$ & $F_{j1}$ & $F_{j2}$ & $F_{j3}$ \\\hline
$\mathbf{F}_1$ & 2.43\% & 1.91\% & 1.73\% \\\hline
$\mathbf{F}_2$ & 0.79\% & 1.87\% & 1.31\% \\\hline
\end{tabular}

\caption{\label{tab:relative error}Relative errors of different operators.}
\end{table}

\end{document}